\documentclass{Interspeech}
\usepackage{multirow}
\usepackage{array}
\usepackage[font=footnotesize]{subcaption}
\usepackage{booktabs}       
\usepackage{graphicx}
\usepackage{color, colortbl}
\usepackage{comment}

\interspeechcameraready

\title{Stack Less, Repeat More: \\ A Block Reusing Approach for Progressive Speech Enhancement}

\author[affiliation={1,*}]{Jangyeon}{Kim}
\author[affiliation={2,*}]{Ui-Hyeop}{Shin}
\author[affiliation={1}]{Jaehyun}{Ko}
\author[affiliation={1,2,{\dagger}}]{Hyung-Min}{Park}

\affiliation{Department of Artificial Intelligence}{Sogang University}{Republic of Korea}
\affiliation{Department of Electronic Engineering}{Sogang University}{Republic of Korea}
\email{ \{jykim97, dmlguq123, jhko, hpark\}@sogang.ac.kr }


\begin{document}

\maketitle

\begin{abstract}
    
This paper presents an efficient speech enhancement (SE) approach that reuses a processing block repeatedly instead of conventional stacking. Rather than increasing the number of blocks for learning deep latent representations, repeating a single block leads to progressive refinement while reducing parameter redundancy. We also minimize domain transformation by keeping an encoder and decoder shallow and reusing a single sequence modeling block.
Experimental results show that the number of processing stages is more critical to performance than the number of blocks with different weights. Also, we observed that the proposed method gradually refines a noisy input within a single block. Furthermore, with the block reuse method, we demonstrate that deepening the encoder and decoder can be redundant for learning deep complex representation.
Therefore, the experimental results confirm that the proposed block reusing enables progressive learning and provides an efficient alternative for SE.



\end{abstract}

\section{Introduction}
\renewcommand{\thefootnote}{}
\footnotetext{*Equal contribution.}
\footnotetext{$^\dagger$Corresponding Author.}
 \renewcommand{\thefootnote}{\arabic{footnote}}

With the rapid development of deep learning, speech enhancement (SE) has been significantly improved as a critical pre-processing step in various applications. In particular, from the conventional real-valued masking in time-frequency (TF) with Short-Time Fourier Transform (STFT), the estimation of complex mask has led to substantial improvement of SE~\cite{dcunet, dccrn}. Furthermore, from U-Net-based architectures~\cite{dcunet, dccrn, cleanunet, frcrn, mfnet}, dual-path modeling in the TF domain further boosted performance in both speech separation (SS)~\cite{tfpsnet, tfgridnet, loco} and SE~\cite{dpt_fsnet, cmgan, mpse-isca, semamba}. More recently, in addition to mask-based approaches, studies based on direct spectral mapping also has been explored, which eliminates the need for masking by outputting enhanced signals either partially~\cite{mpse-isca, semamba} or in full~\cite{tfgridnet, loco}.

Progress in dual-path modeling has been driven by the development of sequence modeling capacity of Transformer-based architectures~\cite{loco, dpt_fsnet, cmgan, mpse-isca}. Transformer-based dual-path models have demonstrated remarkable performance improvements by stacking multiple layers. At the same time, while model performance has continued to improve, there has been a notable shift where model sizes have generally decreased, whereas computational complexity has increased. This trend is evident in the transition from U-Net-based architectures to Transformer-based TF models. It is also reported that the extent of feature processing, influenced by computational complexity, plays a more critical role than the complexity of domain transitions between different representations with a number of parameters and non-linear projections in the model~\cite{scalability}.

Based on this observation, we consider an alternative approach where we reuse a processing block repeatedly instead of stacking multiple processing blocks. These weight sharing across layers have been actively explored in automatic speech recognition~\cite{ShareASR1, ShareASR2} or natural language processing~\cite{ShareNLP1, ShareNLP2}. However, since these tasks often involve transforming input into different domains, sharing weights can lead to performance limitations~\cite{adapter1}. To mitigate this, partial weight sharing~\cite{ShareASR1} or adapter-based mechanisms~\cite{adapter1, adapter2, ShareASR3} have been introduced. On the other hand, in SS, such iterative block reusing technique has been successfully applied~\cite{unfold, tda} and shown to be effective. This is likely because SS operates within the same domain for both input and output. When the input and output domains remain the same, we hypothesize that \textit{progressive refinement} through iterative block reuse can serve as a more efficient alternative to learning deep latent representations using a model with multiple blocks.

Motivated by this, we introduce the block reusing method in SE. Compared to conventional stacking, reusing a block does not result in significant performance degradation in SE. Furthermore, we investigated the impact of the number of stacked and repeated blocks. Empirically, the number of processing stages itself plays a more critical role for performance than the number of blocks with unique weights. Also, our experiments show that training with a repeated block naturally leads a single block to progressively refine the input in contrast to conventional stacked architectures.
Moreover, a deep encoder-decoder structure for learning deeper latent representations can be redundant. To validate our hypothesis, we selected MP-SENet~\cite{mpse-isca} as a baseline which has a convolutional encoder-decoder (CED) architecture that has been widely adopted in recent models~\cite{dcunet, dccrn, mfnet, cmgan}. Then, we minimized domain transformation by keeping CED shallow while increasing the repetition of a single Transformer-based dual-path block in the intermediate layer. Notably, the proposed model retains competitive performance despite the reduction in parameters and computational complexity. This demonstrates that reusing blocks enables progressive feature refinement while significantly reducing parameters, making it an efficient approach for SE tasks without relying on deep latent features learned through a deep encoder and decoder.

\begin{figure}[t]
    \centering
    \subfloat[block stacking]{%
        \includegraphics[width=0.223\textwidth]{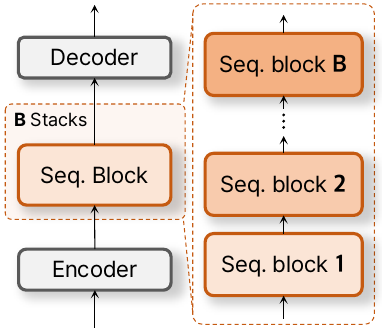}
        \label{fig:stack_repeat_a}
    }\hspace{0mm}
    \subfloat[block repeating]{%
        \includegraphics[width=0.223\textwidth]{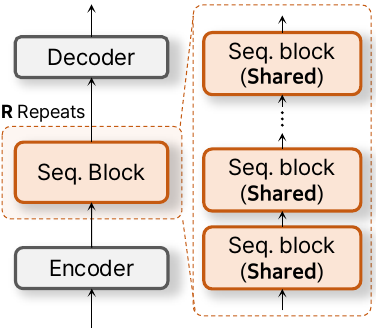}
        \label{fig:stack_repeat_b}
    }
    \vspace{-1mm}
    \caption{Comparison of block (a) stacking and (b) repeating}
    \vspace{-2mm}
    \label{fig:stack_repeat}
\end{figure}

\section{The proposed method}

The SE task fundamentally involves generating an output signal that retains only the desired speech components while remaining in the same domain as the input. Therefore, rather than learning complex representations with multiple blocks, a progressive refinement approach through iterative processing with a single block can serve as an effective strategy.

\subsection{Overview of SE network}

As illustrated in Figure~\ref{fig:stack_repeat}(a), conventional SE methods are often based on the CED architecture with stacked sequence modeling blocks~\cite{cmgan, mpse-isca, semamba}. When a noisy signal is given in the TF domain with $T$ frames and $F$ frequency bins, the noisy input is encoded as a feature representation with the shape of ${C \hspace{-.5mm}\times\hspace{-.5mm} T\hspace{-.5mm} \times\hspace{-1mm} F}$ in TF domain, where $C$ is the feature dimension.
The encoded feature is then processed by a stack of $B$ sequence modeling blocks.
Finally, the decoder transforms the output feature back to its original shape to estimate either the mask value or the direct speech signal.

\subsection{The proposed block reusing method}
Instead of stacking multiple blocks, we propose using a single sequence modeling block repeatedly as shown in Figure~\ref{fig:stack_repeat}(b), which may lead to progressive enhancement without explicit intermediate loss. In particular, this strategy, often referred to as the unfolding method, can have some variants depending on the information fusion scheme between processing stages~\cite{unfold}. 
When using the output of a shared block as input in subsequent iterations, we can simply reuse the output of the previous stage as input to the next stages as a direct connection (DC).
Meanwhile, we can also add initial feature to the each stage output as a summation connection (SC) to encourage leveraging the input feature.
On the other hand, simply reusing the single block can be challenging for more complex noisy inputs. Therefore, we can consider repeating stacks of blocks to enhance modeling capacity of sequence block.

\subsection{Dual-path TF model for SE}
\label{sec:senet}

\begin{figure}[t]
\vspace{-3mm}
\hspace{1mm}
    \centering
    \subfloat[TF-Locoformer]{%
        \includegraphics[width=0.19\textwidth]{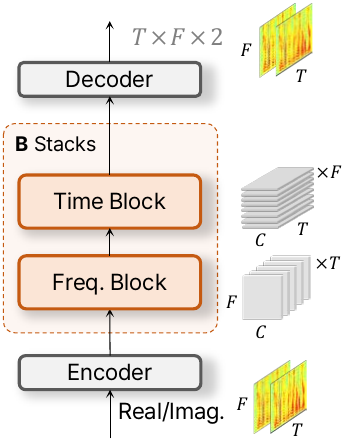}%
        \label{fig:tf_model_a}
    }\hspace{4mm}
    \subfloat[MP-SENet]{%
        \includegraphics[width=0.21\textwidth]{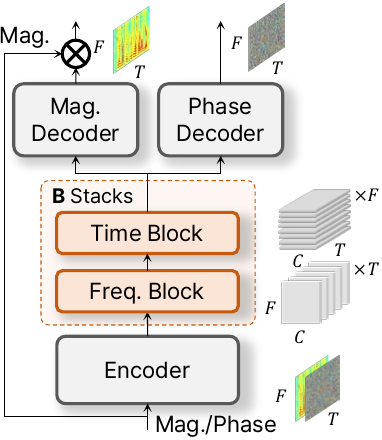}%
        \label{fig:tf_model_b}
    }
    \vspace{-1mm}
    \caption{Two baseline networks using dual-path TF model}
    \vspace{-3mm}
    \label{fig:tf_model}
\end{figure}

To validate and analyze our block reuse method, we selected TF-Locoformer~\cite{loco} and MP-SENet~\cite{mpse-isca} as baseline models for our experiments. Both networks use dual-path TF blocks to model the input features in the TF domain. The dual-path TF blocks alternately process the feature along the time and frequency dimensions to capture temporal and inter-frequency dependencies, respectively, as shown in Figure~\ref{fig:tf_model}.

\subsubsection{TF-Locoformer}
\vspace{-.5mm}
The recently proposed TF-Locoformer~\cite{loco} achieved impressive performance in SS and SE. As in Figure~\ref{fig:tf_model}(a), it follows a CED framework, consisting of only a single Conv2D and Deconv2D layer in the encoder and decoder, respectively. From the real and imaginary (RI) components of a noisy input, TF-Locoformer directly estimates the RI components of the enhanced signal. Also, as a unit module for the dual-path TF block, TF-Locoformer uses a modified feed-forward network to enhance local modeling based on convolutional layers with a Swish-gated linear unit (Conv-SwiGLU) as macaron-style structure in Transformer. Refer to~\cite{loco} for detailed structure.

\subsubsection{MP-SENet}
\vspace{-.5mm}
As depicted in Figure~\ref{fig:tf_model}(b), MP-SENet combines magnitude masking and phase mapping based on the CED architecture, where the encoder consists of a cascade of a convolutional block, a dilated DenseNet~\cite{DenseNet}, and another convolutional block, while the two parallel decoders are composed of a dilated DenseNet followed by a deconvolutional block for magnitude masking and phase mapping, respectively. As a unit for the dual-path TF block, MP-SENet uses Conformer~\cite{conformer} where a convolutional module is incorporated into a macaron-style Transformer to enhance local modeling capacity.

\section{Experimental setups}

\subsection{Datasets and evaluation}

To train and evaluate the proposed method, we used the Interspeech DNS-Challenge 2020 dataset~\cite{dns} for TF-Locoformer and the VoiceBank+DEMAND dataset~\cite{vb} for MP-SENet.
The DNS dataset consists of 500 hours of clean speech data and over 180 hours of noise data.
We generated training data by mixing clean speech with noise at signal-to-noise ratios (SNRs) ranging from -5 to 15 dB.
For evaluation, we used the DNS non-blind test set without reverberation with SNRs ranging from 0 to 25 dB.
Perceptual evaluation of speech quality (PESQ), short-time objective intelligibility (STOI), and scale-invariant signal-to-distortion ratio (SI-SDR) were used as evaluation metrics.
The VoiceBank+DEMAND dataset includes 11,572 training utterances from 28 speakers with SNRs of 0, 5, 10, and 15 dB, and 824 test utterances from 2 speakers with SNRs of 2.5, 7.5, 12.5, and 17.5 dB.
For the VoiceBank+DEMAND dataset, the evaluation metrics include PESQ, segmental SNR (SSNR), and STOI. As additional metrics, CSIG, CBAK, and COVL were employed to assess different aspects of perceptual quality~\cite{cmgan}. In all experiments, the sampling rate was set to 16 kHz. We also compared the parameter size and the number of multiply-accumulate operations (MACs) for 1-second input.

\begin{table}[t]
\renewcommand{\tabcolsep}{5pt}
\def\arraystretch{0.9}
    \centering
    \footnotesize
    \caption{Evaluation on DNS dataset with various configuration of $B$ and $R$ in TF-Locoformer. The best performance is highlighted in \textbf{bold}, and the second-best performance is \underline{underlined}.}
    \resizebox{0.44\textwidth}{!}{
        \begin{tabular}{ccccccccc}
        \toprule
        \multirow{2}{*}{$B$} & \multirow{2}{*}{$R$} & \textbf{Param.} & \textbf{MACs} & \multicolumn{2}{c}{\textbf{PESQ}} & \textbf{STOI} & \textbf{SI-SDR} \\[1pt]
        \multicolumn{2}{c}{} & {(M)} & {(G/s)} & \textbf{-WB} & \textbf{-NB} & (\%) & (dB) \\
        \midrule
        \multicolumn{2}{c}{Noisy} & - & - & 1.58 & 2.45 & 91.52 & 9.07 \\
        \midrule
        1 & 1 & 0.5 & 8.7 & 2.82 & 3.28 & 96.51 & 18.46 \\
        \midrule
        4 & 1 & 1.9 & 34.8 & 3.41 & 3.74 & 98.10 & 20.70 \\
        8 & 1 & 3.7 & 69.6 & 3.47 & 3.79 & 98.26 & 21.07 \\
        12 & 1 & 5.6 & 104.4 & 3.49 & 3.81 & 98.31 & 21.32 \\
        16 & 1 & 7.4 & 139.2 & \textbf{3.55} & \textbf{3.86} & \textbf{98.41} & \textbf{21.67} \\
        \midrule
        1 & 4 & 0.5 & 34.8 & 3.26 & 3.64 & 97.78 & 19.88 \\
        1 & 8 & 0.5 & 69.6 & 3.39 & 3.73 & 98.08 & 20.51 \\
        1 & 12 & 0.5 & 104.4 & 3.43 & 3.76 & 98.18 & 20.91 \\
        1 & 16 & 0.5 & 139.2 & 3.43 & 3.79 & 98.21 & 20.85 \\
        \midrule
        2 & 8 & 0.9 & 139.2 & 3.48 & 3.81 & 98.31 & 21.06 \\
        4 & 4 & 1.9 & 139.2 & \underline{3.52} & \underline{3.83} & \underline{98.38} & \underline{21.32} \\
        8 & 2 & 3.7 & 139.2 & 3.46 & 3.80 & 98.31 & 21.12 \\
        \bottomrule
        \end{tabular}
    }
    \label{tab:integrate_tf}
\end{table}

\begin{figure}[t]
    \centering
    \hspace{-5mm}
    \includegraphics[width=0.42\textwidth]{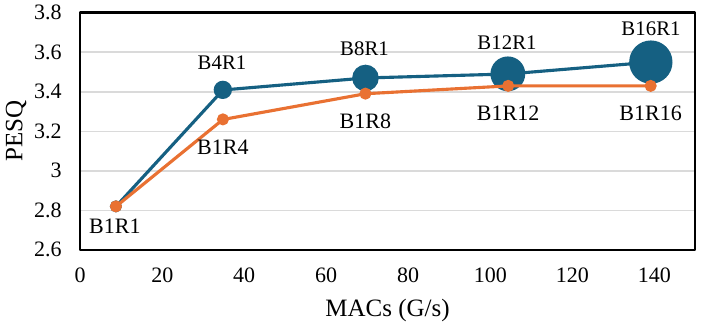}
    \caption{Plot of PESQ-WB vs. MACs on combinations of $B$ and $R$ values from Table~\ref{tab:integrate_tf}. The size of each circle is proportional to the parameter size.}
    \vspace{-2mm}
    \label{fig:macs_vs_pesq}
\end{figure}

\subsection{Training and model configuration}


All models were trained using the AdamW optimizer \cite{adamw} for 100 epochs. 
TF-Locoformer was trained using 4-second segments with a batch size of 1.
STFT was computed using a Hanning window of size 256 and a hop size of 128.
The $C$ was set to 64. A hidden dimension and a kernel size are set to 172 and 3 in the Conv-SwiGLU module~\cite{loco}, respectively. The head of multi-head self-attention is 4.
For training, time-domain $L_1$ loss and TF-domain multi-resolution $L_1$ loss~\cite{loss} was utilized. Note that input normalization and the scaling factor in the loss function from the original work~\cite{loco} were applied only in Section 4.4.
On the other hand, MP-SENet was trained with 2-second segments with a batch size of 2. We followed the same model configuration and training procedure as in original work, including STFT~\cite{mpse-isca}. However, note that the metric discriminator was not used for model training.

\section{Results}
\subsection{Impact of varying the number of blocks and repeats}

Based on TF-Locoformer, we first explored how varying the number of blocks $B$ and repetitions $R$ influences the model's performance on DNS dataset.
Note that all results were derived using the DC method for the information fusion scheme.
In Table~\ref{tab:integrate_tf} and Figure~\ref{fig:macs_vs_pesq}, we observe that as $B$ increases, the results consistently improve along with the corresponding increase in parameters.
Notably, with a single block $B\hspace{-.5mm}=\hspace{-.5mm}1$ that results in only 0.5M parameters, increasing $R$ leads to significant performance improvements across all metrics.
To further explore the trade-off between parameters and model performance through constrained computation costs, we conducted experiments with various combinations of $B$ and $R$ with  $B\hspace{-.8mm}\times\hspace{-.8mm}R$ fixed to 16. Although performance varies depending on the combinations of $B$ and $R$, their differences are not significant because the total number of processing stages $B\hspace{-.5mm}\times\hspace{-.5mm}R$ is constant.
In particular, it is noteworthy that the combination of $B\hspace{-.5mm}=\hspace{-.5mm}4$ and $R\hspace{-.5mm}=\hspace{-.5mm}4$ (denoted as $B4R4$) achieved performance comparable to that of the $B16R1$ configuration, while using about one-fourth of the parameters.
This highlights the effectiveness of the proposed method, demonstrating that it can achieve competitive results with significantly fewer parameters than conventional stacked architectures.

\begin{figure}[t]
    \includegraphics[width=0.46\textwidth]{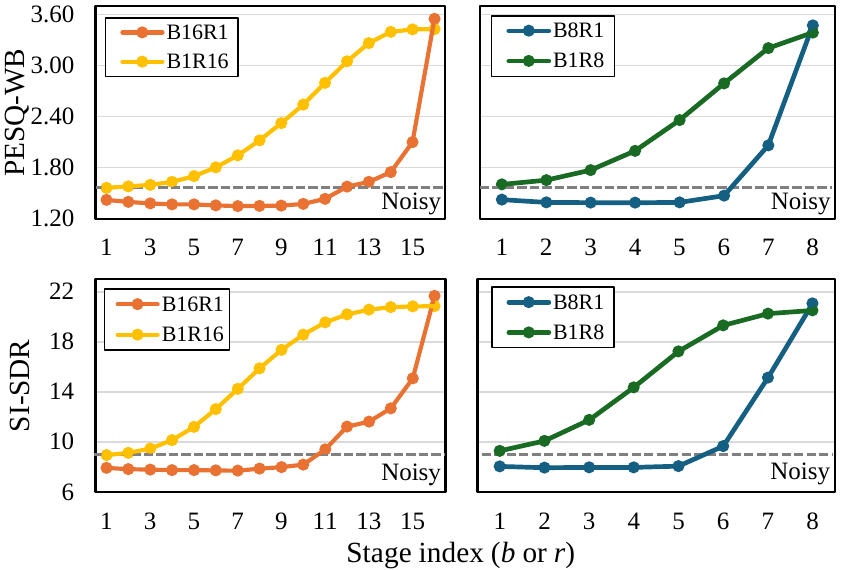}
    \caption{Plots of PESQ-WB and SI-SDR results of intermediate outputs for B16R1, B8R1, B1R16 and B1R8 on DNS dataset.}
    \vspace{-2mm}
    \label{fig:prog_plot}
\end{figure}
\begin{figure}[t]
    \centering
    \subfloat[Sample spectrogram of B16R1]{%
        \includegraphics[width=0.46\textwidth]{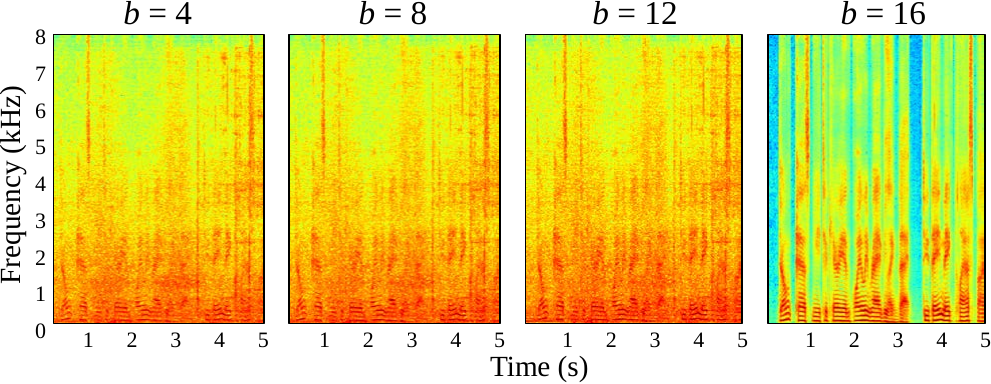}%
    }\\
    \vspace{1mm}\hspace{.5mm}
    \subfloat[Sample spectrogram of B1R16]{%
        \includegraphics[width=0.46\textwidth]{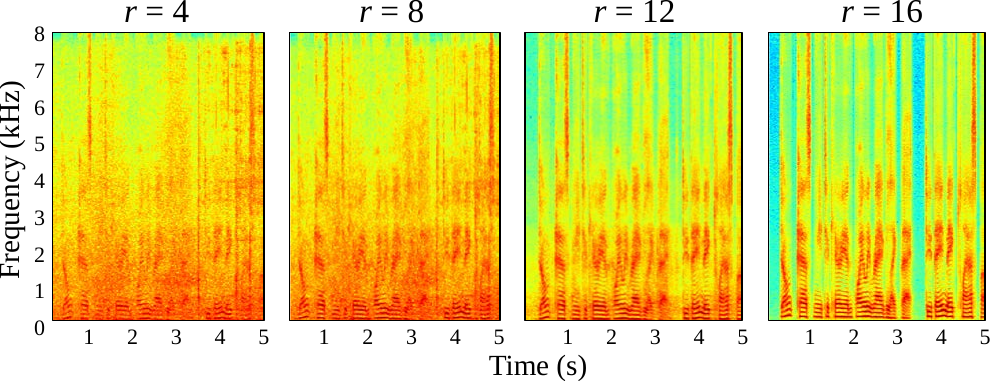}%
    }
    \vspace{-2mm}
    \caption{Spectrograms of a sample utterance from the DNS dataset at different processing stages.}
    \vspace{-2mm}
    \label{fig:spec}
\end{figure}

\subsection{Visualization of progressive enhancement by block reusing}

To analyze the behavior of the network that reuses a single block compared to a network with multiple blocks, we plotted the evaluation results of the intermediate outputs indexed by $1\hspace{-.5mm}\le\hspace{-.5mm}b\hspace{-.5mm}\le\hspace{-.5mm}B$ in $B16R1$ and $B8R1$, and $1\hspace{-.5mm}\le\hspace{-.5mm}r\hspace{-.5mm}\le\hspace{-.5mm}R$ in $B1R16$ and $B1R8$ in Figure~\ref{fig:prog_plot} while using the same decoder in each network. $B1R16$ and $B1R8$ inherently learn to enhance the signal progressively by repeatedly using a single block, starting from a noisy input level. This progressive refinement can also be observed well in the sample spectrogram in Figure~\ref{fig:spec}, which is consistent with the results in Figure~\ref{fig:prog_plot}. 
In contrast, the intermediate outputs in $B16R1$ and $B8R1$ do not show enhancement of the noisy input until $b\hspace{-.5mm}=\hspace{-.5mm}10$ and $b\hspace{-.5mm}=\hspace{-.5mm}5$, respectively. This may be due to the functional partitioning of $B$ blocks into encoder and decoder roles, which could enable the network to learn the deep latent features.

\subsection{Comparison of fusion schemes in block reusing}
\label{subsec:compare_dc_sc}
\begin{table}[t]
\renewcommand{\tabcolsep}{4.5pt}
\def\arraystretch{0.85}
    \centering
    \footnotesize
    \caption{Comparison of fusion schemes in block reusing.}
    \resizebox{0.4\textwidth}{!}{
    \begin{tabular}{lcccccc}
    \toprule
    \multirow{2}{*}{\textbf{System}} & \textbf{Fusion} & \multicolumn{2}{c}{\textbf{PESQ}} & \textbf{STOI} & \textbf{SI-SDR} \\
     & \textbf{scheme} & \textbf{-WB} & \textbf{-NB} & (\%) & (dB) \\
    \midrule
    $B16R1$ & - & \underline{3.55} & \underline{3.86} & \underline{98.41} & \underline{21.67} \\
    \midrule
    \multirow{2}{*}{$B1R16$} & DC & 3.43 & 3.79 & 98.21 & 20.85 \\
     & SC & 3.46 & 3.80 & 98.21 & 21.13 \\
    \midrule
    \multirow{2}{*}{$B4R4$} & DC & 3.52 & 3.83 & 98.38 & 21.32 \\
    & SC & \textbf{3.56} & \textbf{3.87} & \textbf{98.46} & \textbf{21.84} \\
    \bottomrule
    \end{tabular}
    }
    
   \label{tab:compare_dc_sc_ms}
\end{table}

In the network with block reuse, we compared two fusion schemes of DC and SC. Table~\ref{tab:compare_dc_sc_ms} evaluates the case of $B1R16$ and $B4R4$.
The results indicate that the SC method achieves better performance than the DC method by repeatedly adding the input feature, which is consistent with the findings in~\cite{unfold}. The model $B4R4$ with SC method outperformed the model $B16R1$ as the conventional method, with much fewer parameters (7.4M vs. 1.9M). These results highlight the efficiency of the proposed approach in SE.

\subsection{Comparison with existing methods}
\begin{table}[t]
\renewcommand{\tabcolsep}{3.5pt}
\def\arraystretch{0.9}
    \centering
    \footnotesize
    \caption{Comparison of proposed model with previous models on DNS datasets.}
    \vspace{-1mm}
    \resizebox{0.47\textwidth}{!}{%
        \begin{tabular}{lccccccc}
        \toprule
        \multirow{2}{*}{\textbf{System}} & \textbf{Param.} & \textbf{MACs} & \multicolumn{2}{c}{\textbf{PESQ}} & \textbf{STOI} & {\textbf{SI-SDR}\hspace{-3mm}} \\[1pt]
         & {(M)} & {(G/s)} & \textbf{-WB} & \textbf{-NB} & (\%) & (dB) \\
        \midrule
        {Noisy} & & & 1.58 & 2.45 & 91.52 & 9.07 \\
        \midrule
        {FullSubNet~\cite{fsbnet}} & 5.6 & 31.4 & 2.78 & 3.31 & 96.11 & 17.29 \\
        {CTSNet~\cite{ctsnet}} & 4.4 & 5.6 & 2.94 & 3.42 & 96.21 & 16.69 \\
        {TaylorSENet~\cite{taylor}} & 5.4 & 6.2 & 3.22 & 3.59 & 97.36 & 19.15 \\
        {FRCRN~\cite{frcrn}} & 6.9 & 242.0 & 3.23 & 3.60 & 97.69 & 19.78 \\
        {MFNet~\cite{mfnet}} & 6.1 & 6.1 & 3.43 & 3.74 & 97.98 & 20.31 \\
        {USES~\cite{uses}} & 3.1 & 65.3 & 3.46 & - & 98.1 & 21.2 \\
        {TF-Locoformer~\cite{loco}\hspace{-3mm}} & 15.0 & 255.1 & \textbf{3.72} & - & \textbf{98.8} & \textbf{23.3} \\
        \midrule
        {$B1R16$-SC} & 0.5 & 139.2 & 3.53 & \underline{3.85} & 98.37 & 21.71 \\
        {$B4R4$-SC} & 1.9 & 139.2 & \underline{3.64} & \textbf{3.92} & \underline{98.54} & \underline{22.12} \\
        \bottomrule
        \end{tabular}
    }
    \vspace{-1mm}
    \label{tab:compare_prev}
\end{table}
We compared the proposed networks with previous studies on the DNS dataset. For a fair comparison, the proposed networks were trained by applying input normalization and a scaling factor $\ddot{\alpha}$ in the loss function~\cite{loco}. In Table~\ref{tab:compare_prev}, $B1R16$-SC with only 0.5M parameters achieves superior or competitive performance compared to most existing methods, despite its minimum parameters.
Next, we examine the $B4R4$-SC model, which demonstrates the highest parameter efficiency and performance among the proposed methods. 
By a larger margin, this model surpasses most prior works.
Although the proposed models show slightly lower performance compared with TF-Locoformer~\cite{loco}, it is noteworthy that the proposed models maintain much smaller model sizes and about half the computational costs, while still delivering high performance.

\subsection{Investigation of deep encoder and decoder}
\begin{table}[t]
\renewcommand{\tabcolsep}{2.2pt}
\def\arraystretch{0.85}
    \centering
    \footnotesize
    \caption{Evaluation on VoiceBank+DEMAND dataset with various configuration of $K$, $B$, and $R$ in MP-SENet.}
    \vspace{-1mm}
    \resizebox{0.475\textwidth}{!}{
        \begin{tabular}{cccccccccc}
        \toprule
        \multirow{2}{*}{$B$} & \multirow{2}{*}{$R$} & \textbf{Param.} & \textbf{MACs} & \multicolumn{1}{c}{\textbf{PESQ}} & \textbf{STOI} & \textbf{SSNR} & \multirow{2}{*}{\textbf{CSIG}} & \multirow{2}{*}{\textbf{CBAK}} & \multirow{2}{*}{\textbf{COVL}} \\[1pt]
        \multicolumn{2}{c}{} & {(M)} & {(G/s)} & \textbf{-WB} & (\%) & (dB) \\
        \midrule
        \multicolumn{2}{c}{Noisy} & - & - & 1.97 & 91.00 & 1.68 & 3.35 & 2.44 & 2.63 \\
        \midrule
        \multicolumn{10}{c}{\textit{Encoder-decoder with DenseNet of} $K=4$ \textit{(default)}}\\[-1pt]
        \midrule
        4 & 1 & 2.1 & 43.9 & 3.37 & \underline{95.87} & \textbf{10.68} & 4.68 & \underline{3.89} & 4.13 \\
        \midrule
        \multicolumn{10}{c}{\textit{Encoder-decoder with DenseNet of} $K=2$}\\[-1pt]
        \midrule
        4 & 1 & 1.3 & 27.2 & \underline{3.39} & 95.80 & 10.53 & \underline{4.69} & \underline{3.89} & \underline{4.15} \\
        1 & 4 & 0.6 & 27.2 & 3.27 & 94.51 & 10.46 & 4.60 & 3.83 & 4.02 \\
        \midrule
        \multicolumn{10}{c}{\textit{Encoder-decoder with DenseNet of} $K=1$}\\[-1pt]
        \midrule
        4 & 1 & 1.1 & 22.4 & 3.29 & 95.43 & 10.00 & 4.64 & 3.81 & 4.06 \\
        1 & 4 & 0.4 & 22.4 & 3.26 & 95.44 & 10.25 & 4.62 & 3.81 & 4.03 \\
        6 & 1 & 1.5 & 32.0 & \textbf{3.41} & \textbf{96.05} & \underline{10.59} & \textbf{4.74} & \textbf{3.90} & \textbf{4.18} \\
        1 & 6 & 0.4 & 32.0 & 3.36 & 95.65 & 10.21 & 4.68 & 3.86 & 4.13 \\
        \bottomrule
        \end{tabular}
    }
    \vspace{-1mm}
    \label{tab:integrate_mp}
\end{table}

Finally, we examine the impact of varying the depth $K$ of the dilated DenseNet~\cite{DenseNet} in the encoder and decoder of MP-SENet, which is directly related to the degree of domain transformation.
In Table~\ref{tab:integrate_mp}, starting from the default setting of $K\hspace{-.5mm}=\hspace{-.5mm}4$, we evaluated the cases of $K\hspace{-.5mm}=\hspace{-.5mm}2$ and $K\hspace{-.5mm}=\hspace{-.5mm}1$.
When $B\hspace{-.5mm}=\hspace{-.5mm}4$, reducing $K$ significantly decreases both the parameters and computational cost. In particular, compared to $K\hspace{-.5mm}=\hspace{-.5mm}4$, the case of $K\hspace{-.5mm}=\hspace{-.5mm}1$ requires nearly half of both the parameters and computational cost. In spite of the reduced model size and complexity, the performance degradation observed for $K\hspace{-.5mm}=\hspace{-.5mm}1$ remains relatively minor.
More notably, the model with $K\hspace{-.5mm}=\hspace{-.5mm}2$ achieves comparable performance to the case of $K\hspace{-.5mm}=\hspace{-.5mm}4$.

With a shallow encoder and decoder with $K\hspace{-.5mm}=\hspace{-.5mm}1$, the case with $B\hspace{-.5mm}=\hspace{-.5mm}6$, $R\hspace{-.5mm}=\hspace{-.5mm}1$ outperforms the default configuration, achieving the best performance even with fewer parameters and computations. Additionally, the $B\hspace{-.5mm}=\hspace{-.5mm}1$ and $R\hspace{-.5mm}=\hspace{-.5mm}6$ configuration leads to a substantial reduction in parameters, with a only slight performance decrease.
In particular, the model with $B\hspace{-.5mm}=\hspace{-.5mm}1, R\hspace{-.5mm}=\hspace{-.5mm}6$ achieves competitive results compared to the default setting, even with significantly fewer parameters (0.4M). These results consistently support the idea that complex domain transformation with a deep encoder-decoder is not essential, whereas the number of processing iterations plays a more critical role in SE tasks. 

\section{Conclusion}



We presented a progressive SE framework that reuses a processing block efficiently. We confirmed that the number of processing iterations is more critical than the parameter size. Repeating a single block enables progressive refinement, leading to effective performance with fewer parameters. Additionally, we demonstrated that minimizing domain transformation through a shallow CED and block reuse enables progressive feature refinement in SE tasks while significantly reducing parameters without relying on deep latent representations.

However, this work has several limitations that should be addressed in future research. 
First, we focused on denoising tasks in non-reverberant conditions. Future studies should investigate the model's performance in reverberant environments to assess its robustness in real-world scenarios.
Also, we experimented using only dual-path TF model. Further experiments are required on various models to validate the general performance of the proposed approach.
Finally, our study primarily examined non-causal models. Since SE tasks often require real-time applications such as hearing-aids and telecommunication systems, future work should consider causal architectures.


\bibliographystyle{IEEEtran}
\bibliography{mybib}

\end{document}